\documentclass[aps,pra,twocolumn,groupedaddress,amsmath,amssymb,superscriptaddress]{revtex4-1}
\usepackage{graphicx}  
\usepackage{dcolumn}   
\usepackage{bm}        
\usepackage{verbatim}   
\usepackage{upgreek}   
\usepackage{hyperref}
\usepackage{braket}
\usepackage{lineno}
\usepackage[load-configurations=abbreviations, redefine-symbols=true]{siunitx}
\usepackage{xcolor}
\usepackage[normalem]{ulem}

\usepackage[utf8]{inputenc}
\begin{document}
\title{A mid-infrared magneto-optical trap of metastable strontium for an optical lattice clock}
\begin{abstract}

We report on the realization of a magneto-optical trap (MOT) for metastable strontium operating on the \SI{2.92}{\micro\meter} transition between the energy levels $5s5p~^3\mathrm{P}_2$ and $5s4d~^3\mathrm{D}_3$. The strontium atoms are initially captured in a MOT operating on the \SI{461}{\nano\meter} transition between the energy levels $5s^2~^1\mathrm{S}_0$ and $5s5p~^1\mathrm{P}_1$, prior to being transferred into the metastable MOT and cooled to a final temperature of \SI{6}{\micro\kelvin}. Challenges arising from aligning the mid-infrared and \SI{461}{\nano\meter} light are mitigated by employing the same pyramid reflector to realize both MOTs. Finally, the \SI{2.92}{\micro\meter} transition is used to realize a full cooling sequence for an optical lattice clock, in which cold samples of $^{87}$Sr are loaded into a magic-wavelength optical lattice and initialized in a spin-polarized state to allow high-precision spectroscopy of the $5s^2~^1\mathrm{S}_0$ to $5s5p~^3\mathrm{P}_0$ clock transition.

\end{abstract}

\author{R.~Hobson}\thanks{these authors contributed equally to this work}
\email[]{richard.hobson@npl.co.uk}
\affiliation{National Physical Laboratory, Hampton Road, Teddington TW11 0LW, United Kingdom}

\author{W.~Bowden}\thanks{these authors contributed equally to this work}
\email[]{richard.hobson@npl.co.uk}
\affiliation{National Physical Laboratory, Hampton Road, Teddington TW11 0LW, United Kingdom}

\author{A.~Vianello}
\affiliation{National Physical Laboratory, Hampton Road, Teddington TW11 0LW, United Kingdom}
\affiliation{Imperial College London - Department of Physics, London SW7 2BB, United Kingdom}


\author{I.~R.~Hill}
\affiliation{National Physical Laboratory, Hampton Road, Teddington TW11 0LW, United Kingdom}

\author{Patrick Gill}
\affiliation{National Physical Laboratory, Hampton Road, Teddington TW11 0LW, United Kingdom}
\affiliation{Imperial College London - Department of Physics, London SW7 2BB, United Kingdom}
\affiliation{Clarendon Laboratory, Parks Road, Oxford OX1 3PU, United Kingdom}

\maketitle 
\section{Introduction}

\begin{figure}[h]
    \centering
    \includegraphics[width=0.5\textwidth]{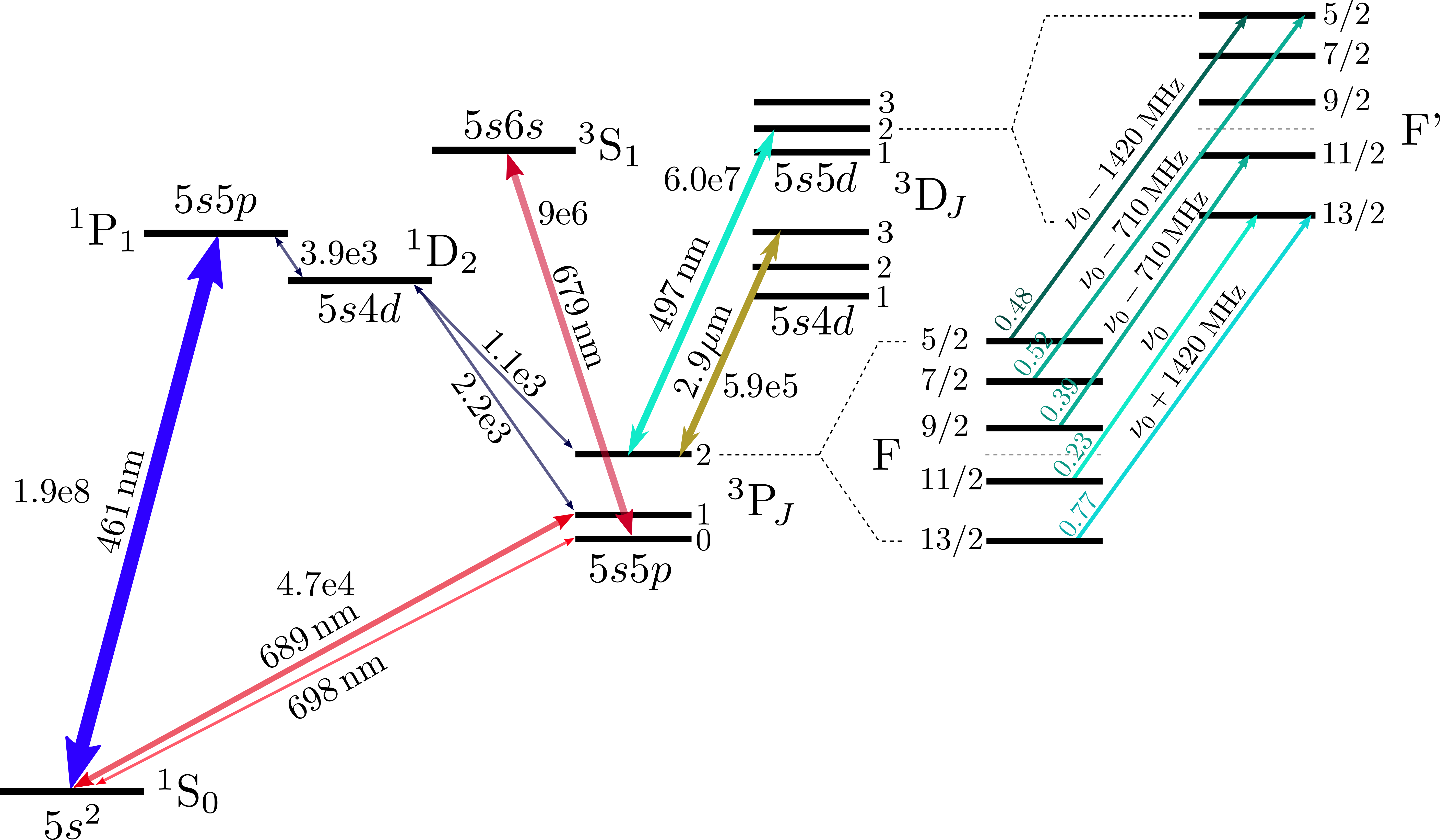}
    \caption{\textit{Left:} Energy level diagram for strontium showing the relevant transitions used in this work. \textit{Right:} In $^{87}$Sr the nuclear spin $I = 9/2$ gives rise to hyperfine structure, as shown for the $5s5p~^3$P$_2$ and $5s5d~^3$D$_2$ states. The transitions between these states is addressed by a modulated repump laser at \SI{497}{\nano\meter} with a carrier frequency $\nu_0$ resonant with the $F = 11/2 \rightarrow F^\prime = 13/2$ transition, and with modulation sidebands addressing transitions from the other hyperfine states. All decay rates are given in units of s$^{-\!1}$.\label{fig:levels}}
\end{figure}

Laser cooling of alkaline earth and alkaline-earth-like atoms is a key tool for realizing our most precise atomic clocks \cite{Ye2008, Nicholson2015, McGrew2018}, with further applications in atom interferometry \cite{Jamison2014, Hu2017, akatsuka2017}, quantum computing and quantum simulation \cite{Daley2008, Gorshkov2009}. 
The ability to cool alkaline earth atoms quickly to a few \si{\micro\kelvin} is especially important for high performance atomic clocks, as the cooling time usually limits the frequency stability of the clock via the Dick effect \cite{Dick1987}. To address this need, efficient methods of laser cooling alkaline earth metals have already been developed in the pioneering work of several groups over the past few decades \cite{Katori1999, Mukaiyama2003, Loftus2004, Sorrentino2006, Stellmer2013}. In this report, we add to this body of work by demonstrating a metastable magneto-optical trap (MOT) for strontium based on the transition between the  $5s5p~^3\mathrm{P}_2$ and $5s4d~^3\mathrm{D}_3$ states separated by \SI{2.92}{\micro\meter} (see Figure \ref{fig:levels}), a cycling transition similar to that used previously to create metastable MOTs for calcium \cite{Grunert2002} and magnesium \cite{Riedmann2012}. We show that the metastable MOT for strontium reaches a final temperature of \SI{6}{\micro\kelvin}, and that the same transition at \SI{2.92}{\micro\meter} can be used to prepare spin-polarized samples of cold $^{87}$Sr atoms in an optical lattice for realizing a high-precision optical clock.

The motivation to pursue a metastable MOT for strontium is that it offers advantages over the typically used \lq red\rq\, MOT based on the $5s^2~^1\mathrm{S}_0 \rightarrow 5s5p~^3\mathrm{P}_1$ transition at \SI{689}{\nano\meter}. The red MOT capitalizes on a narrow transition linewidth of \SI{7.5}{\kilo\hertz} to reach temperatures as low as \SI{250}{\nano\kelvin} \cite{Loftus2004}---close to the limit set by a single \SI{689}{\nano\meter} photon recoil---but the low scatter rate also limits the maximum radiation force to $15g$, making the red MOT relatively shallow and slow to reach equilibrium. 
The red MOT is complicated further 
for $^{87}$Sr because the lack of electronic angular momentum in the $5s^2~^1\mathrm{S}_0$ ground state severely reduces the MOT confinement force unless an additional `stirring' laser is introduced \cite{Mukaiyama2003}. In contrast, the broader transition linewidth of $2\pi\times$\SI{57}{\kilo\hertz} \cite{Safronova2013} in the metastable MOT leads to a higher Doppler limit of \SI{1.3}{\micro\kelvin}, but 
relaxes the requirements for laser frequency stabilization, while the electronic angular momentum of the metastable atoms results in an efficient magneto-optical trapping force for $^{87}$Sr even without the need of an extra stirring laser. Additionally, the long wavelength of the trapping light offers an intriguing possibility to investigate long-range dipolar interactions \cite{Olmos2013} and sets a recoil limited temperature of \SI{12}{\nano\kelvin} which could potentially be accessible via sub-Doppler cooling. A final advantage of the metastable MOT is that it does not strongly perturb the intercombination transitions $^1\mathrm{S}_0 \rightarrow {}^3\mathrm{P}_0$ or $^1\mathrm{S}_0 \rightarrow {}^3\mathrm{P}_1$. This opens the door toward realizing a continuous source of cold atoms for an atom laser \cite{Bennetts2017}, zero-dead-time clock \cite{Schioppo2016}, superradiant laser \cite{Norcia2016,Westergaard2015} or atom interferometer \cite{Hu2017}. 

\section{Laser System and optical setup}
\label{Laser System}

\begin{figure}[tbh]
    \centering
    \includegraphics[width=0.4\textwidth]{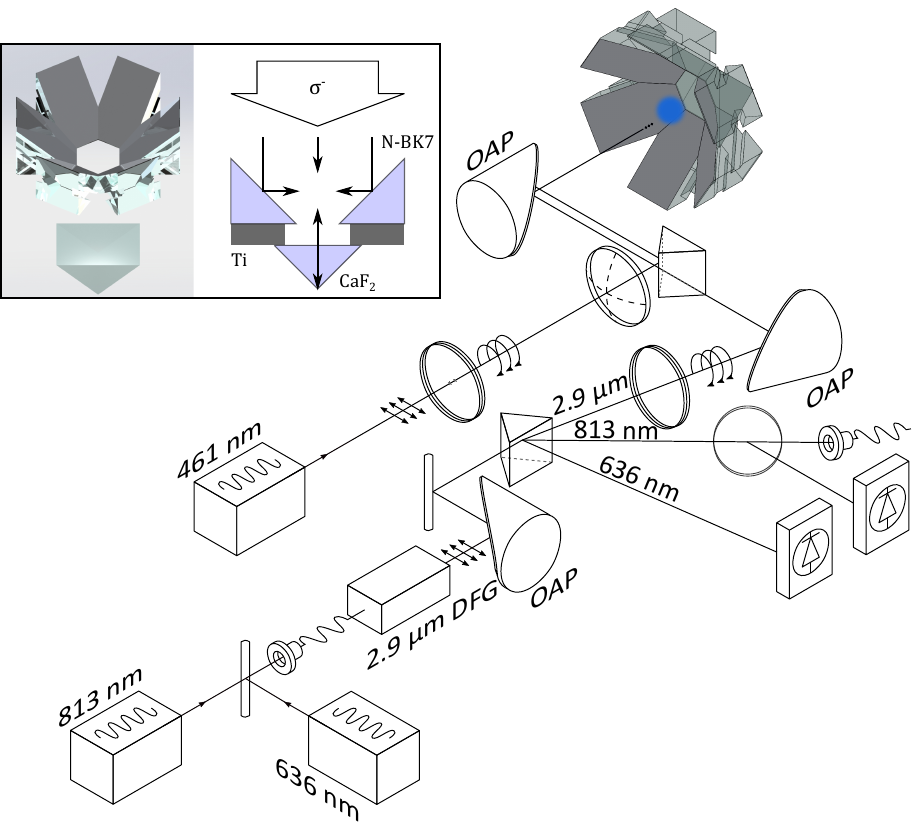}
    \caption{Optical delivery system for the metastable MOT. OAP: Off-axis parabolic mirror. \textit{Top left:} Rendering and cross section of the pyramid MOT optics. A protected silver coating is applied to the hypotenuse of each N-BK7 prism; the CaF$_2$ prism is uncoated. The pyramid MOT assembly measures a total of \SI{33}{\milli\meter} across. \label{fig:pMOT_sketch}}
\end{figure}

The \SI{2.92}{\micro\meter} light needed to operate the metastable MOT is synthesized via difference frequency generation (DFG) from light at \SI{813}{\nano\meter} and \SI{636}{\nano\meter}, and is delivered to the atoms using the optical setup in Figure \ref{fig:pMOT_sketch}. The two optical beams are combined on a dichroic mirror and then coupled into a waveguide embedded in a crystal of periodically-poled lithium niobate \cite{Nishida2005}. As the two optical beams propagate through the waveguide, they interact non-linearly to generate a mid-infrared beam at \SI{2.92}{\micro\meter}. At the output, all three wavelengths of light are collimated on an off-axis parabolic mirror and then separated from each other using a dispersive CaF$_2$ prism cut close to the Brewster angle. The powers in the transmitted beams at \SI{813}{\nano\meter} and \SI{636}{\nano\meter} are actively stabilized on separate photodiodes, allowing feedback to acousto-optic modulators in the optical beam paths for high-precision dynamic control of the \SI{2.92}{\micro\meter} power. The \SI{2.92}{\micro\meter} beam is then expanded using a 1:10 parabolic mirror telescope to a final collimated beam waist $w_0 = \SI{26}{\milli\meter}$ incident on the pyramid MOT. The telescope also provides an opportunity to combine the \SI{2.92}{\micro\meter} beam with the \SI{461}{\nano\meter} MOT beam using a knife-edge mirror placed at the focus. With \SI{125}{\milli\watt} at \SI{813}{\nano\meter} and with up to \SI{25}{\milli\watt} at \SI{636}{\nano\meter} exiting from the waveguide, the DFG module generates up to \SI{700}{\micro\watt} of power at \SI{2.92}{\micro\meter}. Given that the transition saturation intensity is $I_\mathrm{sat} = \SI{0.5}{\micro\watt\per\centi\meter\square}$, a peak intensity up to $140~I_\mathrm{sat}$ can therefore be generated in the center of the MOT beam.

The \SI{636}{\nano\meter} light comes from an extended-cavity diode laser (ECDL) with \SI{120}{\milli\watt} single-mode output power. The \SI{813}{\nano\meter} light is generated using a Ti:Sapphire laser, which also serves as the source of the magic-wavelength optical lattice light used to trap the atoms during spectroscopy of the clock transition---see section \ref{sec:state_prep}. The frequency stability of both lasers is ensured using Pound-Drever-Hall locks \cite{Drever1983} to a shared Fabry-P\'erot cavity \cite{Hill2016}. The tunable cavity is further locked to an ultra-stable, low drift clock laser \cite{Ludlow2007} to transfer long-term frequency stability to the \SI{2.92}{\micro\meter} light. The frequency mismatch between the cavity modes and the $^3$P$_2\!\rightarrow$ $^3$D$_3$ resonance is bridged using an \lq electronic sideband\rq\, configuration, implemented using waveguide EOMs placed between each seed laser and the cavity \cite{Thorpe2008}. As the \SI{813}{\nano\meter} laser must be operated precisely at the magic wavelength to ensure accuracy of the optical clock \cite{Ye2008}, the \SI{2.92}{\micro\meter} cooling light is set to the desired frequency by tuning the offset of the \SI{636}{\nano\meter} laser from the cavity. During preparation of the manuscript, we became aware of similar work using DFG of  \SI{1062}{\nano\meter} and \SI{779}{\nano\meter} light to generate \SI{2.92}{\micro\meter} light which was used to perform an absolute frequency measurement of the same $^3$P$_2\!\rightarrow$ $^3$D$_3$ transition \cite{Hashiguchi2019}.

\section{The metastable MOT sequence}\label{sec:MOTsequence}

The Sr atomic beam is initially slowed using a permanent-magnet Zeeman slower \cite{Hill2014}. Then, all stages of cooling, trapping and clock spectroscopy are implemented inside the pyramid MOT structure depicted in figure \ref{fig:pMOT_sketch}---comprised of six radial silver-coated prism mirrors arranged in a hexagon, plus a CaF$_2$ prism positioned at the apex. Details regarding the prism mirror assembly, together with data illustrating the performance of the first stage \lq blue\rq\, MOT, have been presented in a separate work \cite{Bowden2019}. For the metastable MOT described in this report, the pyramidal retro-reflector has one notable advantage: it greatly simplifies optical alignment and polarization control. In contrast, setting up out-of-vacuum optics for a conventional six beam MOT would be more challenging due to the high cost of detectors and optics in the mid-infrared.

%
The energy level diagram for strontium is shown in figure \ref{fig:levels}. In the first cooling stage, atoms exiting the Zeeman slower are gathered into a \lq blue\rq\, MOT operating at \SI{461}{\nano\meter} on the $5s^2$~$^1$S$_0 \rightarrow 5s5p$~$^1$P$_1$ transition \cite{Kurosu1990}. The transition is not perfectly closed as approximately one in every 50,000 photons scattered causes the atom to be shelved into the metastable 5s5p~$^3\textrm{P}_2$ state.
To recycle atoms in the $^3\textrm{P}_2$ state back into the MOT, a \SI{497}{\nano\meter} repump beam is used. The repump excites the metastable atoms to $5s5d~^3\textrm{D}_2$, which then decays to the ground state via $5s5p~^3\textrm{P}_1$. The \SI{497}{\nano\meter} source is based on a wavemeter-stabilized ECDL operating at \SI{994}{\nano\meter} which is coupled into a waveguide second-harmonic generation module \cite{Nishida2005}. To ensure all five hyperfine sub-levels of the $^3\textrm{P}_2$  state are repumped efficiently, the laser is frequency modulated using a resonant electro-optic modulator (EOM) placed after the second-harmonic generation stage (see Figure \ref{fig:levels}). The laser carrier is tuned to be resonant with the $F= 11/2 \rightarrow F^\prime = 13/2$ transition, while the EOM phase modulation is tuned to \SI{710}{\mega\hertz} with approximately \SI{1.7}{\radian} depth, diverting a significant amount of power into both the first-and second-order sidebands. Additional current modulation is applied to the ECDL at \SI{100}{\kilo\hertz}, which dithers the laser frequency with a deviation of approximately \SI{50}{\mega\hertz} and compensates for any residual frequency detuning between the sidebands and other hyperfine transitions. The modulated \SI{497}{\nano\meter} laser is sufficient by itself to enhance the MOT lifetime by a factor of 30, but an additional laser at \SI{679}{\nano\meter} further improves the lifetime by repumping the small fraction of atoms which leak to the other metastable state, 5s5p $^3\textrm{P}_0$.

\begin{figure}
\centering
    \includegraphics[width=0.4\textwidth]{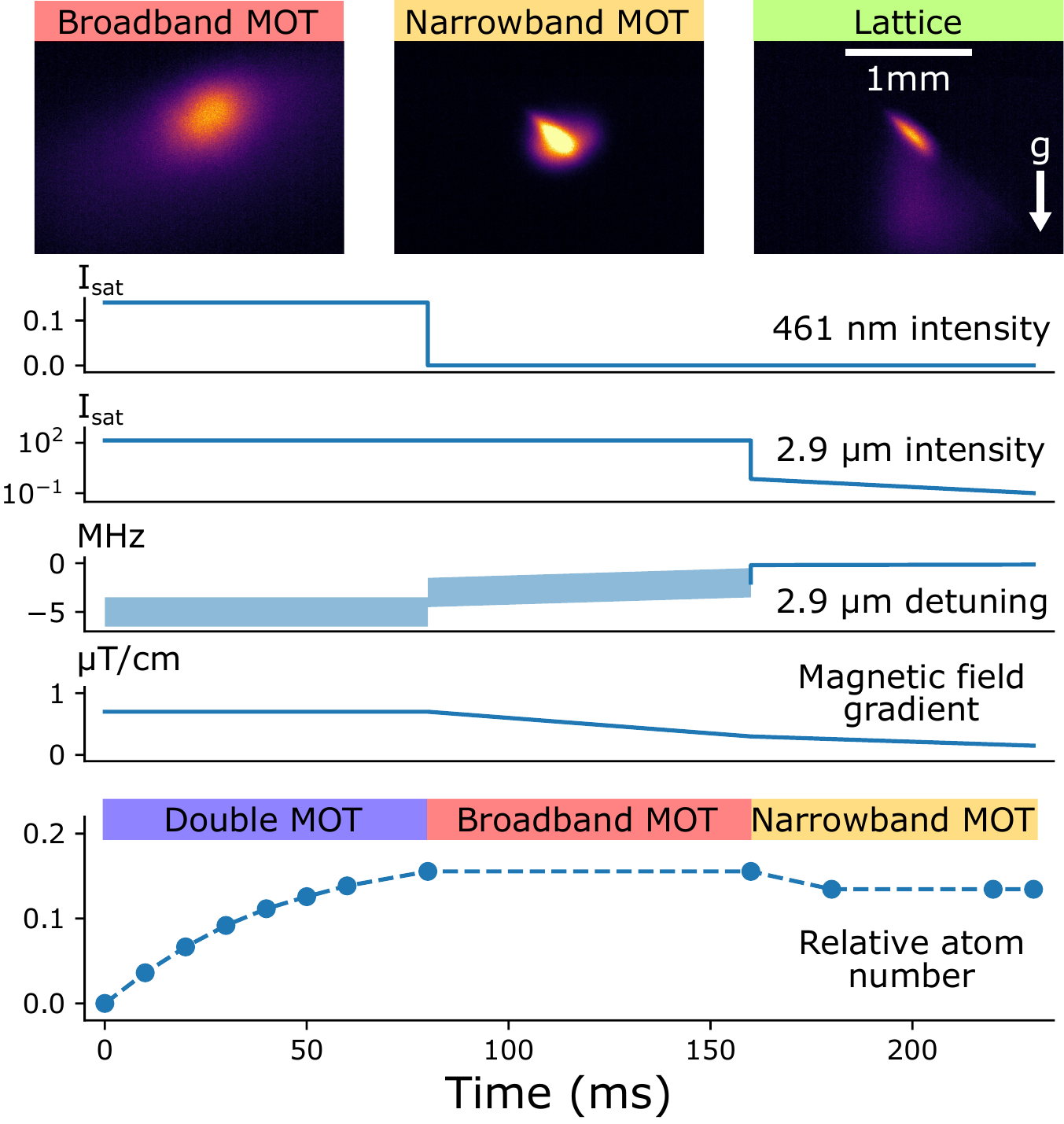}
    \caption{Cooling sequence for the metastable MOT showing the relevant stages: the double MOT which transfers atoms from the blue MOT into the metastable MOT, followed by broadband and narrowband cooling stages using the \SI{2.9}{\micro\meter} light. At each stage the relevant MOT parameters including intensity, laser detuning and field gradient along with the resulting atom number are shown. The relative atom number in the metastable MOT is measured by integrating the florescence signal, and has been scaled so that the number of atoms at the end of the blue MOT is 1. The final narrowband MOT detuning for this dataset is \SI{-113}{\kilo\hertz}. For the double and broadband MOT stages, the laser is frequency modulated as indicated by the detuning being plotted as a band, rather than a line. The top row shows fluorescence images of the MOT at the end of the broadband and narrowband stages, along with a picture of lattice-trapped atoms \SI{25}{\milli\second} after the narrowband MOT light has been switched off. 
    \label{fig:coolingSequence} }
\end{figure}

Once a sufficient number of atoms have been loaded into the blue MOT, the Zeeman slowing beam is switched off and the \SI{2.92}{\micro\meter} light is switched on so that the blue MOT and the metastable MOT can both run concurrently. At the beginning of this \lq double MOT\rq\, stage the \SI{497}{\nano\meter} modulation is switched off so that the $5s5p$ $^3$P$_2$ $(F = 13/2)$ state is no longer repumped, which results in a steady leakage of atoms from the blue MOT via the 5s4d~$^1$D$_2$ state into the metastable MOT operating on the $5s5p$ $^3$P$_2$ $(F = 13/2)$ $\rightarrow$ $5s4d$ $^3$D$_3$ $(F = 15/2)$ transition. The unmodulated \SI{497}{\nano\meter} light remains on during the double MOT so that any atoms which leak into the $5s5p$ $^3$P$_2$ $(F = 11/2)$ state are recycled back into the blue MOT. Under these conditions the metastable MOT loads to half its maximum atom number in \SI{25}{\milli\second} and reaches an asymptotic transfer efficiency of 16\% (see Figure \ref{fig:coolingSequence}). In order for the metastable MOT light to be able to slow and confine the atoms within a sufficiently large region to capture few-\si{\milli\kelvin} atoms efficiently from the blue MOT, the magnetic field gradient is stepped down from \SI{4}{\milli\tesla\per\centi\meter} to \SI{0.7}{\milli\tesla\per\centi\meter} at the beginning of the double MOT, while the \SI{2.92}{\micro\meter} light is operated at maximum intensity of $I_\mathrm{ax} = 140~I_\mathrm{sat}$ in the centre of the axial MOT beam, detuned by \SI{-5}{\mega\hertz} from resonance, and frequency modulated at \SI{400}{\kilo\hertz} with a depth of \SI{1.5}{\mega\hertz}. As well as compensating for inhomogeneous Zeeman shifts, the high intensity and frequency modulation also help to compensate the Doppler broadening resulting from the thermal distribution of atoms exiting the blue MOT. 

Once the transfer into the metastable MOT is complete, the \SI{461}{\nano\meter} light is switched off and the atoms are cooled and compressed in a broadband metastable MOT stage lasting \SI{80}{\milli\second}. The \SI{2.92}{\micro\meter} beam intensity and frequency modulation are kept at the same value as during the double MOT, but the mean detuning is ramped from \SI{-3}{\mega\hertz} to \SI{-2}{\mega\hertz} while the field gradient ramps from \SI{0.7}{\milli\tesla\per\centi\meter} to \SI{0.3}{\milli\tesla\per\centi\meter}. At the end of the broadband stage, the atom cloud has a full width at half maximum of \SI{0.7}{\milli\meter}. To increase the density and reduce the temperature of the cloud further, we implement a final stage of narrowband MOT cooling during which the frequency modulation is switched off. In the narrowband MOT we initially set a relatively large laser detuning and intensity, \SI{-159}{\kilo\hertz} ($1.7\times\Gamma$) and $0.7~I_{\mathrm{sat}}$ respectively, to minimize losses between the broadband and narrowband MOT. Following this, throughout the \SI{70}{\milli\second} duration of the narrowband MOT the magnetic field gradient is ramped linearly from \SI{0.3}{\milli\tesla\per\centi\meter} to \SI{0.15}{\milli\tesla\per\centi\meter}, the intensity is ramped to $0.1~I_\mathrm{sat}$, and the detuning is ramped to a final value in the range \SI{-200}{\kilo\hertz} to \SI{-20}{\kilo\hertz} depending on the desired final size of the cloud (see Figure \ref{fig:TversusD}).

\section{Characterisation of the metastable MOT}

The temperature of the metastable MOT is characterized using a time-of-flight sequence, in which a variable delay is introduced between release of the MOT and the imaging pulse. The Gaussian spatial width of the cloud $\sigma(t)$ as a function of expansion time is related to atomic temperature $T$ using the following expression \cite{Brzozowski2002}:

\begin{equation}
\label{eqn:top}
\sigma(t) = \sqrt{\sigma_0^2 + k_bTt^2/m},
\end{equation}

\begin{figure}
    \centering
    \includegraphics[width=0.5\textwidth]{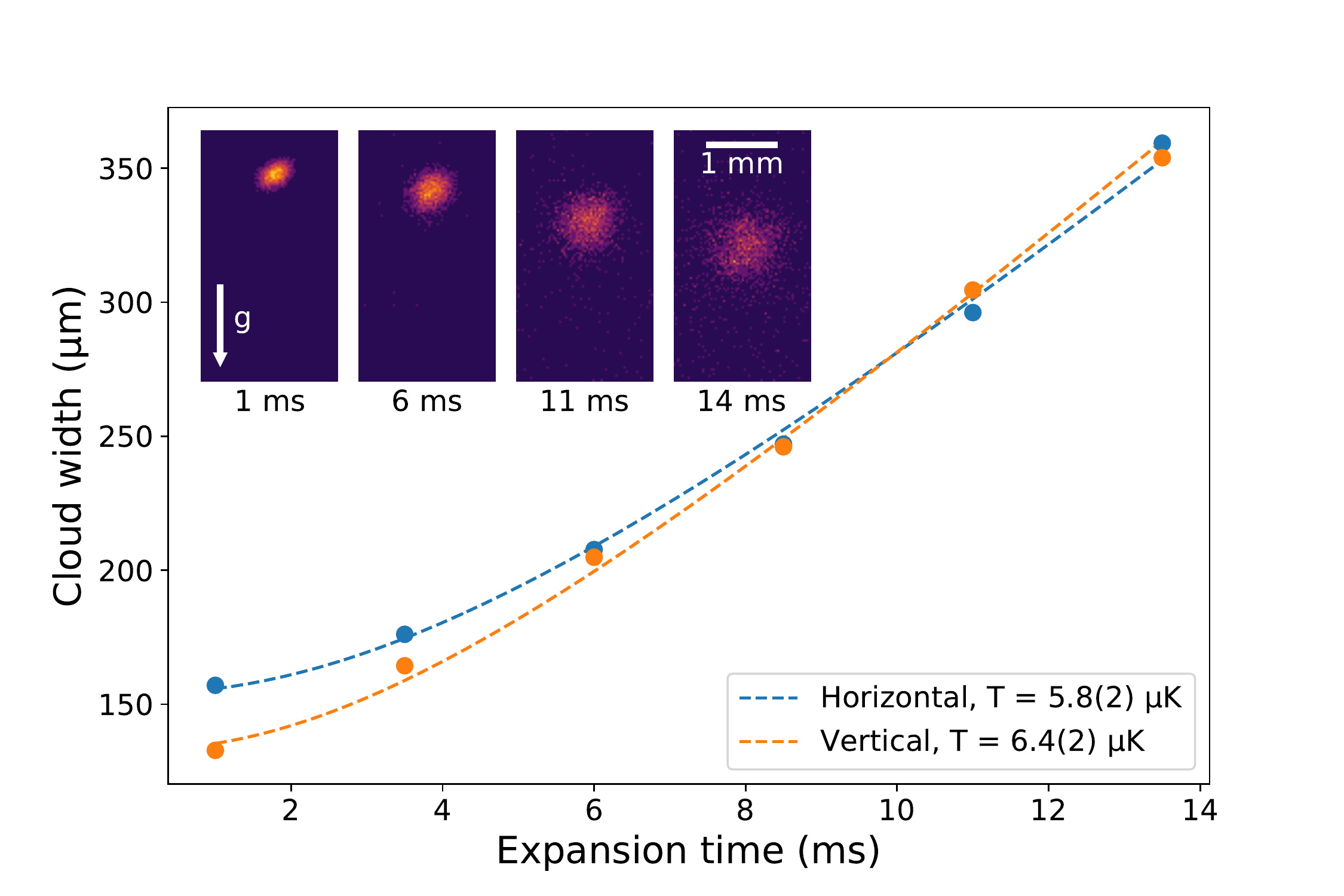}
    \caption{Time-of-flight measurement used to track the expansion rate of the atoms released from the triplet MOT. The expansion rate gives an estimated average temperature of \SI{5.9}{\micro\kelvin}.\label{fig:TOF}}
\end{figure}

\noindent An example time-of-flight measurement is depicted in Figure \ref{fig:TOF} for final MOT detuning and intensity of \SI{-68}{\kilo\hertz} and $0.1~I_{\mathrm{sat}}$ respectively. The images are gathered by the following protocol: after completing the sequence described in section \ref{sec:MOTsequence} the narrowband MOT is held for an additional \SI{70}{\milli\second} in steady state to allow the position and temperature of the atoms to reach equilibrium before the cloud is released and allowed to propagate for a variable time-of-flight between \SI{0}{\milli\second} and \SI{14}{\milli\second}. The atoms are then pumped to the ground state using a \SI{1}{\milli\second} pulse of \SI{497}{\nano\meter} light, and finally imaged using fluorescence from a pulse of resonant \SI{461}{\nano\meter} light lasting \SI{50}{\micro\second}. A 2D Gaussian is then fit to each fluorescence image to extract the cloud size.

\begin{figure}[th]
    \centering
    \includegraphics[width=0.5\textwidth]{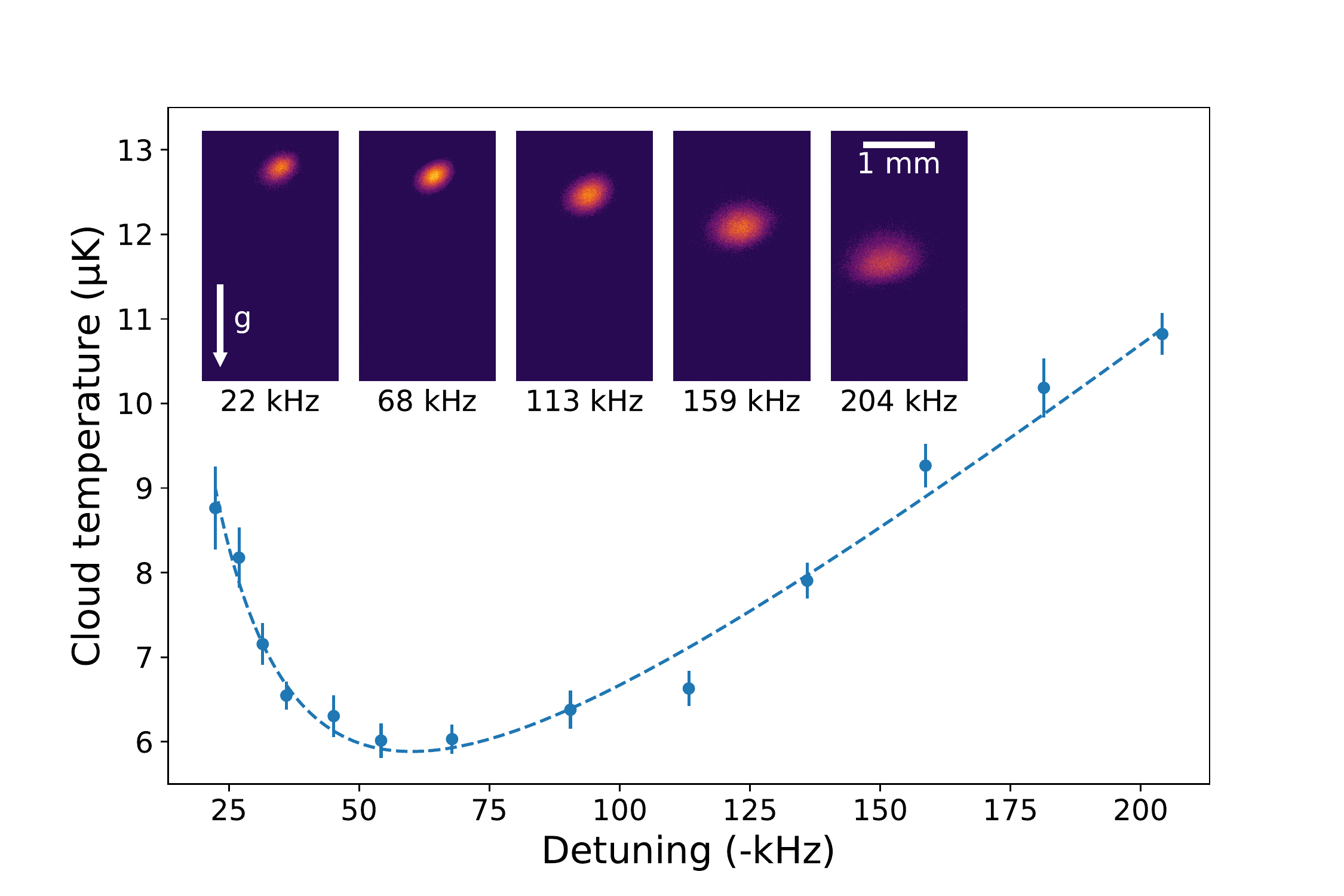}
    \caption{Average temperature of the MOT as measured from time-of-flight expansion. Error bars are estimated from the shot-to-shot noise in the data; they do not include any possible systematic effects. Inset are pictures of the cloud at various detunings. For large detuning, the MOT increases in size and falls under the force of gravity, resting on a shell of maximum force where the Zeeman shift is equal to laser detuning. Similar observations have been made in red MOTs of strontium \cite{Loftus2004}, in both cases indicating that the atom temperature is low enough for the gravitational potential to play a substantial role in the MOT dynamics. \label{fig:TversusD}}
\end{figure}

Repeating this procedure for various final laser detunings $\Delta$, the measured MOT temperature, averaged over the horizontal and vertical directions, is plotted against laser frequency in Figure \ref{fig:TversusD}. According to Doppler cooling theory, the temperature should depend on detuning as \cite{Lett1989}:

\begin{equation}
\label{eqn:temp}
T(\Delta) = \alpha T_{\mathrm{min}}\frac{1+4(\Delta/\Gamma_E)^2}{4|\Delta|/\Gamma_E}.
\end{equation}
where $\Gamma_E = \Gamma\sqrt{1+s_\mathrm{tot}}$ is the power-broadened transition linewidth and $T_\mathrm{min} = \hbar\Gamma_E/2k_B$ is the generalized Doppler cooling limit including the effect of total saturation $s_\mathrm{tot} \approx 6I_\mathrm{ax}/I_\mathrm{sat}$ due to all MOT beams. The overall scaling term $\alpha$ should be unity in the ideal Doppler cooling model, which assumes a two-level atom and no additional heating processes beyond spontaneous emission, but a scaling is included here to match the treatment in \cite{Loftus2004}.


Leaving $\alpha$, $s_\mathrm{tot}$, and the overall detuning as free parameters for the fit curve in Figure \ref{fig:TversusD}, we measure a minimum temperature of $\SI{5.9(6)}{\micro\kelvin}$, corresponding to $\alpha$~=~2.0(1). Furthermore, from the fit we estimate total saturation $s_\mathrm{tot}~=~3.4(3)$, which is factor of two above the measured value based on the beam intensity and trap geometry. The discrepancy in the measured saturation parameter and observation that the temperature did not reduce with intensity, indicates there may be underlying heating effects which we have not fully accounted for.

\section{State preparation for an optical lattice clock}
\label{sec:state_prep}


In this section we outline the additional steps that are needed to use the metastable MOT as the basis for an optical lattice clock---specifically loading the atoms in a magic-wavelength lattice \cite{Ye2008} at \SI{813}{\nano\meter} and optically pumping the sample into a specific Zeeman sub-level for spectroscopy. 

To load atoms into the optical lattice, the lattice light is kept on throughout the final narrowband metastable MOT stage described in section \ref{sec:MOTsequence} so that a significant fraction of the atoms stays trapped after the MOT light is switched off. Using a lattice waist of \SI{100}{\micro\meter} and an axial trapping frequency of \SI{62}{\kilo\hertz} for ground-state atoms, a transfer efficiency of 10\% into the lattice is observed. It is critical that atoms are cooled to the low \si{\micro\kelvin} regime and held in a relatively shallow lattice, so as to constrain their motion while mitigating higher-order shifts from the lattice light \cite{Brown2017,Ushijima2018}.


Before the system can be operated as an optical lattice clock, the atoms in the lattice must be pumped into one of the clock states, $5s^2~^1\mathrm{S}_0$ or $5s5p~^3$P$_0$. Furthermore, to eliminate the 1st-order Zeeman shift of the clock frequency we must average interleaved measurements using atoms prepared in different Zeeman sub-levels with opposite linear magnetic field sensitivities. Typically this is achieved using spin polarized samples, alternating each sequence between the two $M_F= \pm 9/2$ stretched states. In order to prepare the atoms into the $5s^2~^1\mathrm{S}_0$ $M_F= 9/2$, we operate a molasses for \SI{15}{\milli\second} on the \SI{2.9}{\micro\meter} transition in a bias field of \SI{64}{\micro\tesla}, with the laser frequency detuned by \SI{100}{\kilo\hertz} from the red side of the $^3\mathrm{P}_2$~$M_F = +13/2$~$\rightarrow$~$^3\mathrm{D}_3$~$M_F = +15/2$ (i.e. detuned by \SI{-1}{\mega\hertz} from the zero-field atomic resonance). The molasses has little effect on the lattice-trapped atom number or temperature, but ensures that $75\%$ of the atoms end up in the $^1\mathrm{S}_0$ $M_F = + 9/2$ state after a \SI{10}{\milli\second} stage of repumping with \SI{497}{\nano\meter} and \SI{679}{\nano\meter} light. Unfortunately, the molasses cannot directly be used to spin polarize the sample in the $M_F = -9/2$ Zeeman level as this would require a positive frequency detuning which would heat the sample. However, atoms prepared in $^1\mathrm{S}_0$ $M_F = + 9/2$ can be \lq flipped\rq\, to $M_F = - 9/2$ by non-adiabatically switching the magnetic field into the opposite direction in \SI{100}{\micro\second}. Finally, we adiabatically rotate the magnetic field back to the original direction so that the clock spectroscopy can be carried out in the same field on both the $M_F = \pm 9/2$ states. Using a bias field of \SI{64}{\micro\tesla} in magnitude, we observe that a \SI{40}{\milli\second} rotation time is sufficiently long to have a negligible impact on spin polarization fraction.

At the end of all these cooling stages, the spin polarization efficiency is 75\% and the atomic temperature is a large fraction of the trap depth. This has the undesirable effects of limiting the contrast of the clock transition and exacerbating systematic shifts from line-pulling, collisions, and higher-order lattice-atom interactions. To address these problems, a simple state selection protocol is implemented. First, the lattice depth is linearly ramped down to around $\SI{5}{\micro\kelvin}$ in \SI{20}{\milli\second}, held at that depth for \SI{20}{\milli\second}, and then ramped back up to the operating depth of \SI{13}{\micro\kelvin} in \SI{20}{\milli\second}. This \lq spilling\rq\, protocol results in an atom temperature of $\SI{3}{\micro\kelvin}$ as measured via clock spectroscopy of the axial motional sidebands \cite{Blatt2009}, but the reduction in temperature comes at the expense of losing approximately half of the atoms. The \SI{20}{\milli\second} ramp time is optimized experimentally, as the minimum time below which hotter samples are measured after the spilling stage. Next, the spin-polarized Zeeman state is selected by driving a resonant \SI{30}{\milli\second} Rabi $\pi$ pulse on the 5s$^2$ $^1$S$_0$ $M_F = \pm 9/2$ to 5s5p $^3$P$_0$ $M_F^\prime = \pm 9/2$ clock transition in a bias field of \SI{64}{\micro\tesla}, followed by a \SI{5}{\milli\second} flash of \SI{461}{\nano\meter} cooling light which clears out any atoms remaining in the ground state. This \lq clearance pulse\rq\, leaves more than 99\% of the atoms in the single internal state 5s5p $^3$P$_0$ $M_F^\prime = \pm 9/2$, ready for interrogation of the clock transition. We calibrate the final number of spin-polarized atoms using cavity-enhanced non-destructive detection \cite{Hobson2019}, and compare against the fluorescence images of the spin-polarised sample to extract fluorescence signal per atom. With this method we measure a typical final atom number in the lattice of \num{2e4} collected from a blue MOT containing \num{4e6} atoms.

\section{Conclusion and Outlook}
\label{sec:Conclusion}

We have realized a mid-infrared MOT for metastable strontium, observing a temperature of \SI{6}{\micro\kelvin} in the low-intensity limit. We use the mid-infrared transition to prepare $^{87}$Sr atoms in a spin polarized state in a magic-wavelength optical lattice for high-precision spectroscopy of the 5s$^2$ $^1$S$_0$ to 5s5p $^3$P$_0$ clock transition. The metastable MOT results in comparable loading time and atom number as alternative cooling methods for optical lattice clocks \cite{Falke2014,LeTargat2006}, and achieves a lower temperature than metastable MOTs previously realized in calcium \cite{Grunert2002} and magnesium \cite{Riedmann2012} by an order of magnitude.

Further optimization of the metastable MOT sequence, aimed at increasing atom number or reducing the cooling time, could potentially improve the performance of the optical lattice clock by reducing the Dick-effect frequency instability \cite{Dick1987}. For example, better transfer into the metastable MOT might be achieved if more optical power were available at \SI{2.92}{\micro\meter}: the double MOT could then be operated at a higher intensity and larger laser frequency modulation depth, thereby increasing the capture velocity and volume of the metastable MOT. For the broadband metastable MOT stage, replacing the sinusoidal modulation with a sawtooth-wave adiabatic-passage (SWAP) cooling protocol could increase phase space density more quickly \cite{Snigirev2019}. Finally, improved transfer efficiency into the optical lattice could be facilitated by the addition of a \SI{497}{\nano\meter} \lq drain\rq\, laser overlapped with the lattice to selectively repump atoms into the ground state as they enter the capture volume---this method has already been successfully implemented as a means to load a dipole trap from a metastable calcium MOT \cite{Yang2007}.

Future work could pursue sub-Doppler cooling on the \SI{2.92}{\micro\meter} transition, and discover how far metastable atoms can be cooled towards the recoil limit of \SI{12}{\nano\kelvin}. Further, one could explore schemes to cool atoms on a continuous basis as proposed in \cite{katori2001}. Such a scheme would be compatible with various quantum sensors reliant on the narrow $^1\mathrm{S}_0 \rightarrow {}^3\mathrm{P}_0$ or $^1\mathrm{S}_0 \rightarrow {}^3\mathrm{P}_1$ transitions given the fact that these transitions are not strongly perturbed by the \SI{2.92}{\micro\meter} light.

\section{Acknowledgments}
\label{sec:Acknowledgments}

This work was financially supported by the UK Department for Business, Energy and Industrial Strategy as part of the National Measurement System Programme; and by the European Metrology Programme for Innovation and Research (EMPIR) project 15SIB03-OC18. This project has received funding from the EMPIR programme co-financed by the Participating States and from the European Union’s Horizon 2020 research and innovation programme.

\bibliography{myBib}

\end{document}